\documentclass[11pt, letter]{article}
\usepackage[margin=2.5cm]{geometry}
\usepackage{amsmath, amssymb, amsthm}
\usepackage{booktabs}
\usepackage{hyperref}
\usepackage{cleveref}
\usepackage{epsfig}
\usepackage{graphicx}

\newtheorem{theorem}{Theorem}[section]
\newtheorem{lemma}[theorem]{Lemma}

\let\eps\varepsilon
\newcommand{\PP}{\mathcal{P}}
\newcommand{\R}{\mathbb{R}}

\newcommand{\Between}{\operatorname{Between}}
\newcommand{\Within}{\operatorname{Within}}

\DeclareMathAlphabet{\mathui}{T1}{cmr}{m}{ui}
\def\dim{\mathui{d}}

\title{A Linear-Size Well-Separated Pair Decomposition \\for Unit Disk Graphs}
\author{Jie Gao\thanks{Department of Computer Science, Rutgers University, Piscataway, NJ 08854. Email: {\texttt jg1555@rutgers.edu} 
}}
\date{}

\begin{document}

\maketitle

\begin{abstract}
Given a metric space $(X, d)$ a $c$-well separated pair decomposition (WSPD) is a collection of `dumbbell' pairs of subsets $(A, B)$, $A, B\subseteq X$, where the distance between the two sets is at least $c$ times the diameter of each set and every pair of points in $X$ is covered by one such pair. It is known that for a weighted unit disk graph metric of $n$ points, there is a WSPD of size $O(n\log n)$. But it is open whether this bound can be improved to $O(n)$. In this note, we answer this question affirmatively and provide an algorithm that yields a linear-size WSPD for unit disk graphs. 
\end{abstract}

\section{Introduction}

Given a metric $(X, d)$, a pair of subsets $(A, B)$ with $A\subseteq X$, $B\subseteq X$ is called a \emph{$c$-well-separated pair}, $c>1$,  if the distance between them is at least $c$ times the diameter of each pair. Here, we use $d(A, B)$ to refer to the distance between two sets $A, B\subseteq X$, $d(A, B)=\min_{a\in A, b\in B} \pi(a, b)$, and $D(A)$ as the diameter of a set 
$A\subseteq X$, $D(A)=\max_{a\in A, a'\in A} d(a, a')$. 
Now, a \emph{$c$-well-separated pair decomposition ($c$-WSPD)} is a set of $c$-well-separated pairs $\PP=\{(A_i, B_i)\}$ that cover every pair of points in $X$. That is,  $\forall p \neq q\in X$, $\exists (A_i, B_i)\in \PP$ such that $p\in A_i$ and $q\in B_i$ or vice versa, and $\forall (A_i, B_i)\in \PP$, $$d(A_i, B_i)\geq c\cdot \max\{D(A_i), D(B_i)\}.$$

We consider the metric $(X, d)$ to be the shortest-path metric of a weighted unit disk graph. In particular, consider $X$ to be a set of $n$ points in $\R^{\dim}$. An edge connects two points if their Euclidean distance is at most $1$. The edge is also given a weight equal to the Euclidean distance. The metric $d(p, q)$ measures the shortest path distance between two points $p, q$ in this graph $G$. It is shown in~\cite{gao03well} that for any $c> 1$ there is a $c$-well separated pair decomposition of the unit disk graph metric of size $O(n\log n)$ when $\dim=2$, and of size $O(n^{2-2/\dim})$ when $\dim\geq 3$. The size of the WSPD for dimension $3$ or higher is tight -- there is an example where any $c$-WSPD has to be of size $\Omega(n^{2-2/\dim})$. But it has been open since~\cite{gao03well}\footnote{Recently the work by Har-Peled et. al. ~\cite{harpeled25well} improved the dependency on $c=1/\eps$ in the running time from $O(1/\eps^4n\log n)$ to $O(1/\eps^2n\log n)$, but the dependency on $n$ stays asymptotically the same. }  whether one can find a linear-size WSPD for the unit disk graph in 2D, or if one can show a stronger lower bound of $\Omega(n\log n)$.  

\paragraph{New Results.}
In this note, we present a new algorithm that produces a $c$-WSPD for any $c>1$ of size $O(n)$, settling this question. The main result is the following theorem. 
\begin{theorem}\label{thm:WSPD}
For a unit disk graph of $n$ points in the plane, there is a $c$-well-separated pair decomposition of size $O(n)$ for any $c>1$.
\end{theorem}

In this document, we first show that the algorithm in~\cite{gao03well} may produce a WSPD of size $\Omega(n\log n)$. Thus, the analysis of the algorithm in~\cite{gao03well} is actually tight in terms of $n$.  The algorithm  in~\cite{gao03well}  takes a bounded-degree spanning tree and recursively performs a balanced partitioning. It turns out if the tree is not carefully chosen, the subtrees or clusters in this hierarchical decomposition may be very close to each other, generating a super-linear number of well-separated pairs.

The improvement to a linear-size WSPD uses a hierarchical net-tree, where the centers at level $i$ are required to be at distance at least $2^i$ apart. To bound the total size, the analysis uses a randomized padded decomposition~\cite{klein93excluded} and a potential function argument to balance the number of pairs across different levels. Note that the padded decomposition is used only for the analysis and the algorithm uses a deterministic net-tree. The running time of the new algorithm is $O(n^2 \log^2n/\log \log n)$, dominated by the time to find all-pairs shortest paths of a unit disk graph~\cite{brewer2026improved}. Note that the algorithm by Gao and Zhang~\cite{gao03well} has running time $O(n\log n)$, as it does not require finding the all-pairs shortest path distances.  Instead, it uses the Euclidean distance as a lower bound of the graph distance and the number of points minus one as an upper bound on the diameter of a connected subgraph.  

Last, we note that Har-Peled et. al. ~\cite{harpeled25well} had an \emph{output-sensitive} algorithm for WSPD of a given metric. If a metric admits an $O(f(n))$ WSPD, the algorithm will be able to find one. Thus, applying Har-Peled's algorithm to the unit disk graph metric would also produce a linear-size WSPD -- since we now know its existence. 

\paragraph{Related Work} Well-separated pair decomposition was initially proposed for points in $\R^{\dim}$ by Callahan and Kosaraju~\cite{CallahanKosaraju1995} -- any $n$ points in $\R^{\dim}$ admits a $c$-WSPD of size $O(n)$. Later, it is shown that a  linear-size WSPD exists for metrics of bounded doubling dimension~\cite{HarPeledMendel2006}. A unit disk graph metric does not have bounded doubling dimension. 
A $1/\eps$-WSPD of size $O(\lambda^2\eps^{-4}n\log n)$ is shown for a $\lambda$-low-density graph~\cite{GudmundssonW2026wspd}. And a linear-size WSPD is shown for $c$-packed graphs~\cite{Deryckere2025}.
Finally, Buchin et. al.~\cite{buchin26small} considered the problem of minimizing the size of a WSPD for a given input point set. 

\paragraph{AI Disclosure} The main technique idea -- the potential function in the proof -- was discovered by ChatGPT Astra. This note is entirely written by the author. 

\section{$\Omega(n\log n)$ Size for the Gao-Zhang Algorithm}

We briefly review the algorithm in~\cite{gao03well}. In this discussion, we only consider a point set of constant density -- the maximum number of points inside a unit disk is $O(1)$. To build a WSPD, we first construct a spanning tree $T$ of the unit disk graph $G$ of maximum degree at most $6$ and then recursively build a balanced decomposition -- on a tree of maximum degree at most $\beta$, there is an edge whose removal partitions the tree into two parts of size at least a fraction of $1/(\beta+1)$. The decomposition can be represented by a binary tree $T'$ in which the root represents the entire tree $T$ and the two children represent the connected subtrees obtained by removing an edge from $T$. Each node $v$ in the tree $T'$ represents a connected subtree $T(v)$ of $T$. Denote by $S(v)$ the set of vertices in $T(v)$.  One vertex of $S(v)$ is chosen as the representative $\sigma(v)$ of $T(v)$. 

The algorithm for constructing a $c$-WSPD in~\cite{gao03well}  is as follows. Start with a pair $(S(r), S(r))$ in a queue, where $r$ is the root of $T'$, and thus $S(r)$ is the set of all points in $X$. Proceed until the queue is empty. In each step, remove the first pair $(S(v_1), S(v_2))$ from the queue and consider the following cases:
\begin{itemize}
\item If $\|\sigma(v_1)-\sigma(v_2)\| \geq (c+2)\cdot \max (|S(v_1)|-1, |S(v_2)|-1)$, insert the pair to the well separated pair decomposition. Here $\|\sigma(v_1)-\sigma(v_2)\|$ is the Euclidean distance between $\sigma(v_1)$ and $\sigma(v_2)$.
\item Otherwise, if $|S(v_1)|= |S(v_2)| =1$, they contain the same point and we discard the pair. Without loss of generality let's assume $|S(v_1)|\geq |S(v_2)|$ and that $u_1,u_2$ are two children of $v_1$, we add to the queue two pairs $(S(u_1),S(v_2))$ and $(S(u_2),S(v_2))$.
\end{itemize}

\begin{figure}[t]
   \centering
    \includegraphics[width=0.47\textwidth]{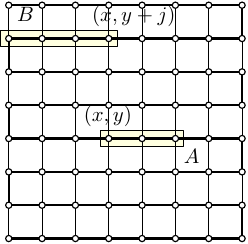}
    \caption{A WSPD of size $\Omega(n\log n)$ by the algorithm in~\cite{gao03well}.}
\label{fig:grid} 
\end{figure}

We show that the algorithm on the following example produces $\Omega(n\log n)$ pairs. Consider $n$ as a power of $2$ and a unit disk graph as a unit grid of $\sqrt{n}\times \sqrt{n}$. All the edges are precisely the grid edges. We also consider a spanning tree $T$ as a path that zigzags from the bottom up. See~\Cref{fig:grid}. The hierarchical decomposition of the tree $T$ iteratively bisects the path. Take $c>2$.

We first argue that any well-separated pair $(A, B)$ produced by the algorithm in~\cite{gao03well} cannot have a set of diameter greater than $\sqrt{n}$. Indeed, the graph diameter is $2\sqrt{n}$. By the design of the hierarchical partition, each well-separated pair $(A, B)$ produced by the algorithm must have all vertices in $A$  (and in $B$) in the same row.

Now consider two points of coordinate $(x, y)$ and $(x, y+j)$ with $x \in [\sqrt{n}]$, $j\in  [\sqrt{n}/2]$ and $1\leq y\leq \sqrt{n}-j$. Suppose they are covered by a $c$-well-separated pair $(A, B)$ and the representative points in the two sets $A$ and $B$ are $\sigma(A)$ and $\sigma(B)$, respectively. Now we know by construction, 
$$d(\sigma(A),\sigma(B)) \geq \|\sigma(A)-\sigma(B)\| \geq (c+2)\cdot \max (|A|-1, |B|-1)$$
and $$d(\sigma(A),\sigma(B)) \leq |A|-1+j +|B|-1$$
Take $m=\max(|A|, |B|)$, we have $m\leq j/c+1\leq 2j$. 
Thus, fix $y$ and $j$, for all pairs $(x, y)$ and $(x, y+j)$ on the two rows, we need at least $\sqrt{n}/(2j)$ well-separated pairs to cover them. 
There are $\sqrt{n}-j$ pairs of rows that are exactly $j$ vertically apart. Thus, the number of well-separated pairs is at least $$\sum_{j=1}^{\sqrt{n}/2} (\sqrt{n}-j) \cdot \frac{\sqrt{n}}{2j}=\Omega(n\log n).$$

\section{Linear-Size WSPD}

We first assume that the points have bounded density; as in~\cite {gao03well}, the algorithm can be extended to the general case. To get the linear size, we will need a different algorithm. In particular, we build a hierarchical decomposition using a graph-metric net-hierarchy. In this section, we will first present the algorithm and then bound the size of the WSPD. 

\subsection{Algorithm}\label{subsec:algo}

Consider a unit disk graph with bounded density. We build a nested net hierarchy at scales $r_i=2^i$ for integers $i\geq 0$.
At level $-1$ define a set $N_{-1}=X$ with all the points. Now we build $N_i\subseteq N_{i-1}$ in a greedy manner. Initialize $N_{i}=\emptyset$. Scan the points in $N_{i-1}$. If a point $p$ is within distance $r_i$ from an already selected point $q$ in $N_i$, assign $p$'s parent as $q$. Break ties arbitrarily. If not, add $p$ to $N_i$. The points in $N_i$ have pairwise graph distance at least $r_i$, and every point in $N_{i-1}$ is within distance $r_i$ from at least one point in $N_i$. 
The net hierarchy has at most $J=O(\log n)$ levels. 

We define the cluster $C(p)$ for a point $p\in N_i$ as the descendants in the hierarchy. The clusters of points in $N_i$ partition $X$. The clusters of different scales are nested. The cluster of a point $p\in N_{i-1}$ is a subset of the cluster of the parent $q\in N_i$ of $p$. 
Any point $x$ in a cluster $C(p)$ with $p\in N_i$ is at distance at most $\sum_{j=0}^{i} 2^j< 2^{i+1}=2r_i$ from $p$. Thus, the graph diameter of a cluster $C(p)$ is at most $4r_i$.

Now we use the net-hierarchy to build a $c$-WSPD. We have two procedures: $\Within(p)$ builds the well-separated pairs covering pairs within $C(p)$, and $\Between(p, q)$, for $p, q\in N_i$, builds the well-separated pairs covering pairs with one from $C(p)$ and the other one from $C(q)$. We start with $\Within(r)$, where $r$ is the root of the net-hierarchy. 

The procedure $\Within(p)$ will quit if $C(p)=\{p\}$ is a singleton cluster. Otherwise, we recursively call $\Within(q)$ for each child $q$ of $p$, and call $\Between(q, w)$ for each unordered pair of children $q, w$ of $p$.

In the procedure $\Between(p, q)$, if $C(p)$ and $C(q)$ are singletons, or if $d(p, q)\geq 4(c+1)  2^i$, we add the pair $(C(p), C(q))$ to the output WSPD.  Otherwise, we call $\Between(a, b)$ for each child $a$ of $p$ and each child $b$ of $q$.

To see why the produced pairs are $c$-well-separated, recall that any point in a cluster $C(p)$ (or $C(q)$) is within distance $2r_i$ from $p$ (or $q$), with $p, q\in N_i$. By the triangle inequality,
$$d(C(p), C(q))\geq d(p, q)-4r_i\geq c \cdot 4r_i\geq c \cdot \max(D(C(p), D(C(q))).$$
The coverage of the produced well-separated pairs is immediate. 

To implement the algorithm, we first run all-pairs shortest paths for the unit disk graph $G$ in time $O(n^2\log^2 n/\log\log n)$~\cite{brewer2026improved}. The recursive call takes time at most $O(n^2\log n)$.

\subsection{Size of WSPD}

We now bound the size of the WSPD by the above algorithm. The main Lemma we prove is the following.

\begin{lemma}\label{lem:bound}
For a weighted unit disk graph with bounded density, at scale $r_i=2^i$, let $N_i$ be any set of vertices that are pairwise more than $r_i$-apart and $E_i=\{\{u,v\}\subseteq N_i:d(u,v)\le Kr_i \}$ for a constant $K$, then $\sum_{i>i_0} |E_i|=O(K^2n)$, for a sufficiently large constant $i_0$. 
\end{lemma}

This Lemma is enough to prove our main claim. In particular, every well-separated pair we produced at each level $i$ is charged to some pair of points $p, q\in N_i$. By the recursive algorithm, their parents $p', q'\in N_{i+1}$ either collide ($p'=q'$) or fail to be well-separated at level $i+1$, in which case, $d(p', q')\leq 4(c+1)2^{i+1}$. Thus $$d(p, q)\leq d(p, p')+d(p', q')+d(q', q)\leq (8c+12)\cdot 2^{i}.$$ 
Thus, each well-separated pair generated by the algorithm can be charged to a pair in $E_i$ in~Lemma \ref{lem:bound}.
The total number of well-separated pairs produced at level at most $i_0$ for a constant $i_0$ is $O(n)$ by the bounded density assumption. 

The proof of this lemma requires the use of padded decomposition of weighted planar graphs. In particular, a weighted unit-disk graph has a planar spanner of constant stretch~\cite{gao2001geometric}. That is, there is a subgraph $H$ of the weighted unit disk graph $G$ such that $d_H(p, q)\leq \alpha \cdot d_G(p, q)$, for a constant $\alpha$. Here $d_H$ refers to the shortest path distance in graph $H$. 

Now consider the planar graph $H$, by~\cite{klein93excluded},
%~\cite{Filtser2019StrongDiameter} 
there is an absolute constant $b\ge1$ such that, for every $\Delta>0$, there is a random partition $\mathcal Q_\Delta$ of $X$ satisfying 
$D_H(C)\le\Delta$\footnote{Follow-up work on padded decomposition for planar graphs (e.g., ~\cite{Filtser2019StrongDiameter}) can produce stronger results where the clusters have small \emph{strong} diameter (using shortest path distance within a cluster). We only need the weak diameter (shortest path distance in graph $H$).} for every cluster $C\in\mathcal Q_\Delta$,
and for any $v\in X$ and $t\ge 0$, 
$$\Pr\!\left[B_H(v,t)\not\subseteq\mathcal Q_\Delta(v)\right]
\le b\,\frac{t}{\Delta}$$
Here $\mathcal Q_\Delta(v)$ denotes the cluster containing $v$ and $B_H(v, r)$ the set of points within distance $r$ from $v$ in graph $H$. 

By the spanner property, we have $B_{G}(v,t)\subseteq B_{H}(v,\alpha t)$. Also $D_G(C) \leq D_H(C) \leq \Delta$. Therefore, the partition $\mathcal Q_\Delta$ of $X$ satisfies $D_G(C)\le\Delta$  and for any $v\in X$ and $t\ge 0$, 
$$\Pr\!\left[B_G(v,t)\not\subseteq\mathcal Q_\Delta(v)\right]
\le b'\,\frac{t}{\Delta},\, b'= \alpha b$$

We will skip the subscript $G$ from now on, and all distances and diameter are in the graph $G$. Now, we are ready to prove the Lemma.
\begin{proof}
We build a nested padded hierarchy on a unit disk graph $G$. For each $0\leq j \leq J=\log n$, we independently sample a partition $\mathcal Q_j$ with $\Delta=2^j$. $\mathcal Q_J=\{X\}$.

For any $0\leq j \leq J$, denote by $\mathcal Q_j(v)$ the cluster in $\mathcal Q_j$ that contains vertex $v$. Define the common intersection $R_j(v)=\cap_{k=j}^J \mathcal Q_j(v)$. If $R_j(v)$ has multiple connected components, split the components into their own clusters. This provides a nested partition $\mathcal P_j$. $\mathcal P_J=\{X\}$. These partitions have the following property: each cluster is connected and has diameter at most $2^j$; $\mathcal P_j$ refines $\mathcal P_{j+1}$ -- one cluster in $\mathcal P_{j+1}$  is decomposed into several clusters in $\mathcal P_{j}$; each cluster in $\mathcal P_j$ is in a cluster of $\mathcal Q_j$. Last, a ball $B(v,t)$ induces a connected subgraph. Consequently, if the ball is contained in every $\mathcal Q_k(v)$, $k\ge j$, then it is contained in the connected component $\mathcal P_j(v)$. By a union bound, we have
\begin{equation}\label{eq:ball}
\Pr[B(v,t)\not\subseteq\mathcal P_j(v)]
\le \sum_{k=j}^{J-1}
   \Pr[B(v,t)\not\subseteq\mathcal Q_k(v)]\\
\le \sum_{k=j}^{\infty}b'\,\frac{t}{2^k}\\
=2b'\,\frac{t}{2^j}=2b'\,\frac{t}{r_j}.
\end{equation}

For the setup of the Lemma, we will relate the pairs in  $E_i$ to the partition $\mathcal P_i$. Consider a pair $\{p,q\}$ in $E_i$. Call the pair \emph{good} if all three events hold:
\[
\begin{aligned}
B(p, \frac{r_i}{16b'})&\subseteq\mathcal P_i(p),\\
B(q,\frac{r_i}{16b'})&\subseteq\mathcal P_i(q),\\
B(p,Kr_i)&\subseteq\mathcal P_{i+h}(p), \, h=\left\lceil\log_2(16b' K)\right\rceil,
\end{aligned}
\]
By~\Cref{eq:ball}, the first two events fail with probability at most $1/8$, and the third fails with probability at most $\frac{2b' K}{2^h}\leq 1/8$. By the union bound, the probability that $(p, q)$ is good is at least $5/8$. Let $Y_i$ be the expected number of good pairs at level $i$. $Y_i \geq |E_i| \cdot 5/8$. 
In the following, we will bound $\sum_{i\geq i_0} Y_i$ to be $O(n)$. This will be enough for the lemma. 

A good pair $\{p, q\}$ has the following property. Since $d(p, q)\leq Kr_i$ and the ball at $p$ of radius $Kr_i$ stays in a cluster $\mathcal P_{i+h}(p)$, $p, q$ both stay in this cluster. Further, $d(p, q)>r_i$ and clusters in $\mathcal P_i$ have diameter at most $r_i$; $ p$ and $ q$ stay in different clusters in $\mathcal P_i$. Last, since the ball at $p$ of radius $r_i/(16b')$ stays in cluster $\mathcal P_i(p)$, the cluster $\mathcal P_i(p)$ has diameter at least $r_i/(16b')-1\geq r_i/(32b')$, if $2^i\geq (32b')$. To see why, follow the shortest path from $p$ to $q$ and stop at the last vertex $z$ right before exceeding distance $r_i/(16b')$ from $p$; $z$ stays in the cluster  $\mathcal P_i(p)$. Since the diameter of the cluster $\mathcal P_i(p)$ is at least $r_i/(32b')$, it has at least $r_i/(32b')$ vertices.

Now fix a cluster $U\in\mathcal P_{i+h}$. Let
$U_1,\ldots,U_k $ be all the descendant clusters of $U$ in $\mathcal P_i$ and let $t$ be the number of these clusters with diameter at least $r_i/(32b')$. The number of good pairs that can be generated from the clusters $\{U_j\}$ is at most ${t \choose 2}$. To bound this value, we use a potential function argument.

For any connected vertex set $C$, we define a potential function $$\Phi(C)=2(|C|-1)-D(C)$$
Since $C$ is connected and each edge has weight at most $1$, $D(C)\leq |C|-1$. Thus $0\leq \Phi(C)\leq 2(|C|-1)$.
Now suppose a cluster $C$ is partitioned into $k$ connected pieces 
$C=C_1\sqcup\cdots\sqcup C_k$. Consider the increase of the potential function 
\begin{equation}\label{eq:gamma}
\Gamma(C)=\Phi(C)-\sum_{i=1}^k \Phi(C_i)=2(k-1)+\sum_{i=1}^kD(C_i)-D(C).
\end{equation}
Recall that $C$ and $C_i$ are all connected subgraphs. If we shrink $C_i$ to a supernode, the subgraph on $C$ has diameter at most $k-1$. Thus $D(C)\le \sum_{i=1}^kD(C_i)+(k-1)$.
This means, 
\begin{equation}\label{eq:gamma0}
\Gamma(C) \geq k-1\geq 0
\end{equation}
 
The increase of the potential function provides an upper bound on $k\geq t$. Further, if we count all good pairs with different values of $i$, the sum of the differences of the potential function allows telescoping, which provides the final $O(n)$ bound. We explain the details next.
 
Now we are ready to bound the value ${t \choose 2}$. Apply \Cref{eq:gamma} for our case of $U$ and its decomposition into $U_1, \cdots, U_k$.  we have 
\begin{equation}\label{eq:gammaU}
 \Gamma(U)\geq 2(k-1)+r_i \cdot \frac t{16b'} - 2^h r_i=2(k-1)+r_i \cdot \left( \frac t{16b'}  -2^h\right)
\end{equation}
Now we do a case study of $t$. 
If $t/(16b')\geq 2\cdot 2^h$, by \Cref{eq:gammaU} we have $\Gamma(U)\geq  r_i \cdot t/(32b')$. Or, $r_i t\leq \Gamma(U) (32b')$. Since the point set has bounded density and $U$ has radius at most $2^{i+h}$, we have $|U|\leq A 2^{2(i+h)}$ for some constant $A\geq 1$. $U$ has $t$ subsets, each of at least $r_i/(32b')$ vertices. Thus $|U|\geq t \cdot 2^i/(32b')$. Combining the two inequalities, we have $t\leq A 2^{2h} (32b')r_i$. Altogether, we have $${t \choose 2}\leq \frac{t^2}{2} \leq A 2^{2h} (16b')r_i t\leq 2A 2^{2h} (16b')^2 \Gamma(U)$$
If $t/(16b')< 2\cdot 2^h$, apply~\Cref{eq:gamma0}, we have $${t \choose 2}= \frac t2 (t-1) \leq  \frac t2 (k-1)\leq 16b' 2^h \Gamma(U)$$
Therefore, in both cases, we have ${t\choose 2}\leq B\cdot \Gamma(U)$ for a constant $B=2A 2^{2h} (16b')^2 =O(K^2)$.

To bound the total number of good pairs at level $i$, $Y_i$, we now sum over all clusters in level $i+h$. Define $F_i=\sum_{C\in \mathcal P_i} \Phi(C)$. 
$$Y_i\leq B \sum_{U\in \mathcal P_{i+h}} \Gamma(U)=B\cdot (F_{i+h}-F_i)$$
Now, $$\sum_{i\geq i_0} Y_i \leq B\sum_{i=0}^J (F_{i+h}-F_i)$$

Recall that the sum of the potential function increases as the level $i$ goes up. $F_i\leq F_{i+1}$. $\mathcal P_J=\{X\}$ and $F_J=\Phi(X)=2n-1-D(X)\leq 2n$. 
Thus, $$\sum_{i\geq i_0} Y_i\leq B h \cdot F_J= O(K^2n)$$
This finishes the Lemma. 
\end{proof}
%Recall that $\mathcal P_j$ is a refinement of $\mathcal P_{j+1}$, thus, the sum of the potential function for all clusters in $\mathcal P_j$ is no greater than the sum of the potential of all clusters in $\mathcal P_{j+1}$. 

To complete \Cref{thm:WSPD}, we use the same idea as in Section 4.2 of~\cite{gao03well} to handle a general set of points, thus dropping the bounded density assumption.

\bibliographystyle{alphaurl}
\bibliography{wspd}
\end{document}